\documentclass[3p,times,procedia]{elsarticle}
\usepackage{nupha_ecrc}

\volume{00}

\firstpage{1}

\journalname{Nuclear Physics A}

\runauth{}

\jid{nupha}

\jnltitlelogo{Nuclear Physics A}

\usepackage{amssymb}

\usepackage[figuresright]{rotating}

\begin{document}

\begin{frontmatter}



\dochead{XXVIIth International Conference on Ultrarelativistic Nucleus-Nucleus Collisions\\ (Quark Matter 2018)}

\title{New measures of longitudinal decorrelation of harmonic flow}


\author{Piotr Bo{\.z}ek}
\address{AGH University of Science and Technology,
 Faculty of Physics and Applied Computer Science,
 aleja Mickiewicza 30, 30-059 Krakow, Poland}

\author{Wojciech Broniowski}

\address{The H. Niewodnicza\'nski Institute of Nuclear Physics, Polish Academy of Sciences, 31-342 Cracow, Poland}
\address{Institute of Physics, Jan Kochanowski University, 25-406 Kielce, Poland}

\begin{abstract}
Harmonic flow in relativistic heavy-ion collisions is observed in a broad range
of rapidities, and  the flow at different rapidities is correlated. However, 
fluctuations lead to a small decorrelation of the  harmonic flow magnitudes and 
flow angles  at different rapidities. 
Using a hydrodynamic model with Glauber Monte Carlo initial conditions we show 
that the flow angle decorrelation strongly depends on the flow magnitude in the 
event. We propose observables to measure this effect 
in experiment. 
\end{abstract}


\end{frontmatter}


\section{Introduction}

The expansion of the fireball created in relativistic heavy-ion collisions 
generates a collective flow of matter. Azimuthal asymmetry of 
the initial fireball generates harmonic components in 
the azimuthal distribution of emitted particles. The most important observables
are the elliptic ($v_2$) and triangular ($v_3$)  flow coefficients.
Without fluctuations in the longitudinal direction, the initial shape of the 
fireball in the 
transverse plane is  similar at different rapidities. This means that the
  pattern of azimuthally asymmetric  collective flow repeats itself
 in a similar way at different rapidities as well.

\begin{figure}
 \begin{center}
 \includegraphics[width=0.35\textwidth]{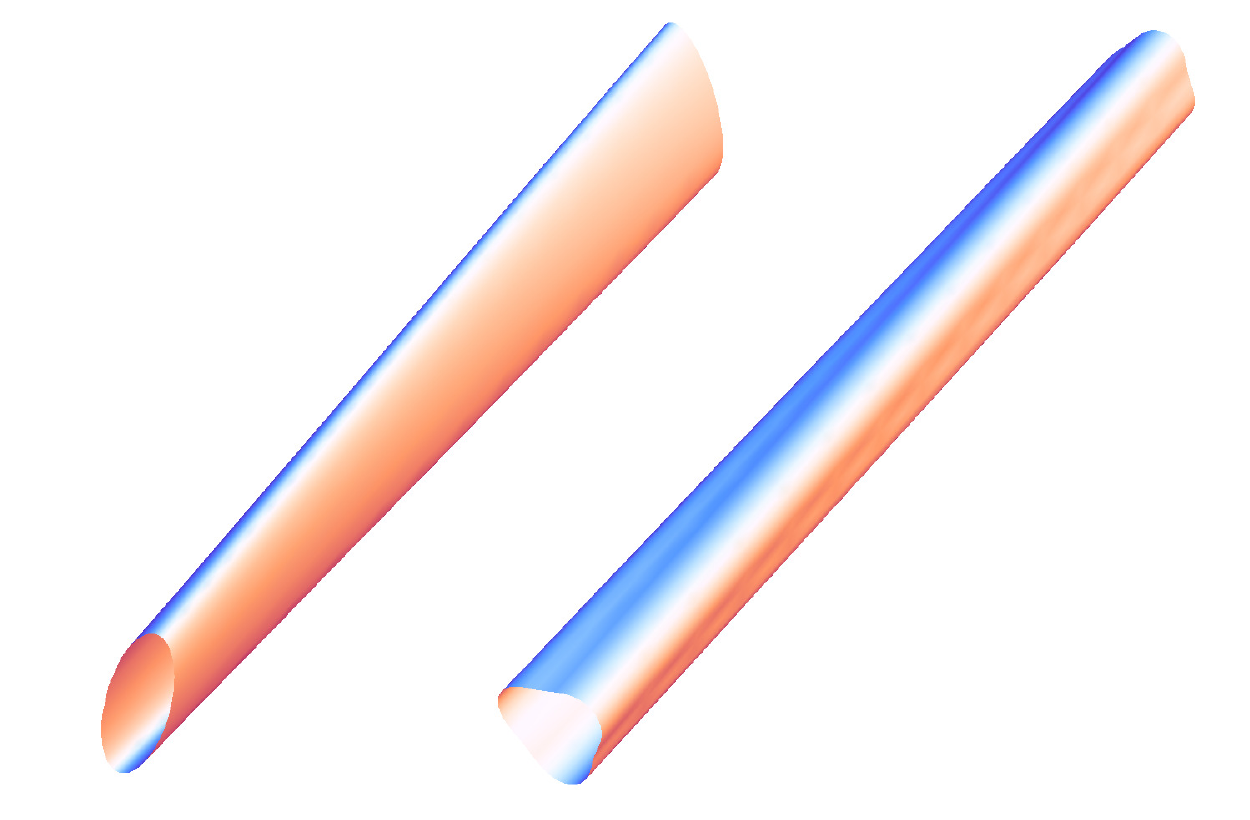}~~~~~~~~~~~~~~~~
 \includegraphics[width=0.4\textwidth]{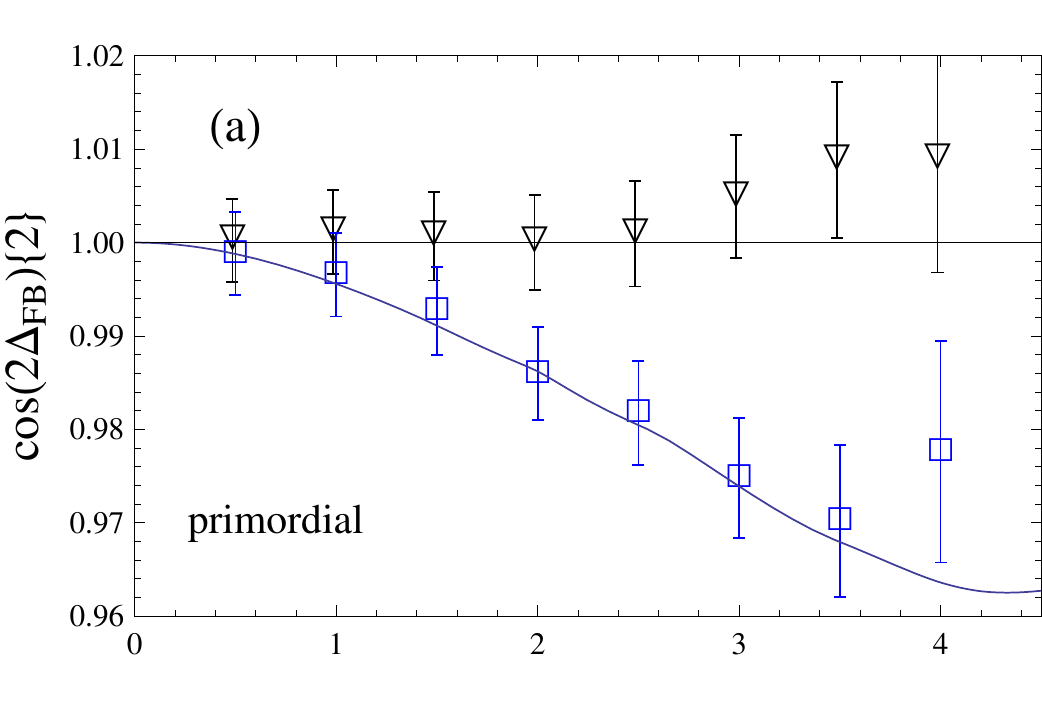}
 \caption{(left panel) Illustration of the torque of the orientation of the initial fireball. The effect is shown 
 for the elliptic and triangular deformation. (right panel) The decorrelation coefficient (Eq. \ref{eq:dc}) 
 of the elliptic flow  from the hydrodynamic model (from \cite{Bozek:2010vz}).}
 \label{fig:old}
 \end{center}
\end{figure}

Event by event fluctuations break this symmetry of the flow 
orientation~\cite{Bozek:2010vz}. This is expected, as event by event  
distributions of the forward- and backward-going participant nucleons are 
different.
 Such fluctuations in the initial state cause a torque on the initial fireball 
(Fig. \ref{fig:old}, left panel). 
The torque angle  in the orientation of the fireball asymmetry
 fluctuates from even to event.   In each event the  flow angle for a given harmonic changes slightly with rapidity. 
Obviously, this effect cannot be observed 
on event by event basis. On the other hand, the average correlation between 
the flow vectors,
\begin{equation}
q_n(\eta)=\frac{1}{m}\sum_{k=1}^{m} e^{i n \phi_k} \equiv v_n(\eta) e^{i n \Psi_n(\eta)},
\end{equation}
measured in two pseudorapidity intervals at  $\pm \eta$,
\begin{equation}
r_n( \eta)= \frac{\langle q_n(\eta) q_n^\star(\eta ) \rangle}{\sqrt{\langle q_n(\eta) q_n^\star( \eta ) \rangle\langle q_n(\eta) q_n^\star(\eta ) \rangle}},
\label{eq:dc}
\end{equation}
can be measured.

The coefficient $r_n(\eta)$ is smaller than $1$ and decreases as  $ \eta$ increases (Fig. \ref{fig:old}, right panel). 
Unfortunately, this quantity is strongly influenced by non-flow correlations~\cite{Bozek:2010vz}.
The CMS Collaboration proposed to use a correlator defined using $3$ pseudorapidity bins,
\begin{equation} 
 r_{n|n;1}(\eta)=\frac{\langle q_n(-\eta)q_n^\star(\eta_{\rm ref})\rangle}{\langle  q_n(\eta)q_n^\star(\eta_{\rm ref}) \rangle} ,
\label{eq:dvn}
\end{equation}
which reduces non-flow contributions \cite{Khachatryan:2015oea}. The decorrelation of  flow (factorization breaking) 
in rapidity   has been confirmed experimentally using the $3$ bin correlator.

\section{Magnitude and flow angle decorrelation}

The correlator defined in Eq. (\ref{eq:dvn}) combines 
two types of harmonic flow decorrelation~\cite{Jia:2014vja}. The first one is the decorrelation of the flow 
magnitude in separate pseudorapidity intervals. This effect means that in an ensemble 
of events a strong flow in one interval does not always come together with 
a strong flow in the other interval. The second effect  measured by
 (\ref{eq:dvn}) is the cosine of the torque angle, i.e. of the 
the decorrelation of the flow angles in the two intervals,
$\Delta \Psi_n = \Psi_n(\eta)-\Psi_n(\eta_{ref})$.
The two effects can be estimated separately using additional correlators of harmonic flows in $3$ or $4$ pseudorapidity intervals~\cite{Jia:2017kdq}. 

The $4$ bin correlator
\begin{equation}
R_n(\eta)=\frac{<q_n(-\eta_{ref})q^\star_n(\eta)q_n(-\eta)q_n^\star(\eta_{ref})>}
{<q_n(-\eta_{ref})q^\star_n(-\eta)q_n(\eta)q_n^\star(\eta_{ref})> }\simeq 1 -2 F^{\rm tor}_{n,2} \eta
\end{equation}
 measures the flow component, whereas the $3$ bin correlator
\begin{equation}
r_{n,2}(\eta)=\frac{<q_n(-\eta)^2q_n^\star(\eta_{ref})^2>}{<q_n(\eta)^2q_n^\star(\eta_{ref})^2>}\simeq 1- 2 F^{\rm asy}_{n,2} \eta -2 F^{\rm tor}_{n,2}\eta
\end{equation}
measures  the torque angle and magnitude decorrelations combined. Measurements by the ATLAS 
Collaboration \cite{Aaboud:2017tql} have shown that the magnitude of the two components for the  decorrelation is similar, $F^{\rm tor}_{n,2}\simeq F^{\rm asy}_{n,2}$.

\section{Torque angle dependence on flow magnitude}

\begin{figure}[h]
 \begin{center}
 \includegraphics[width=0.4\textwidth]{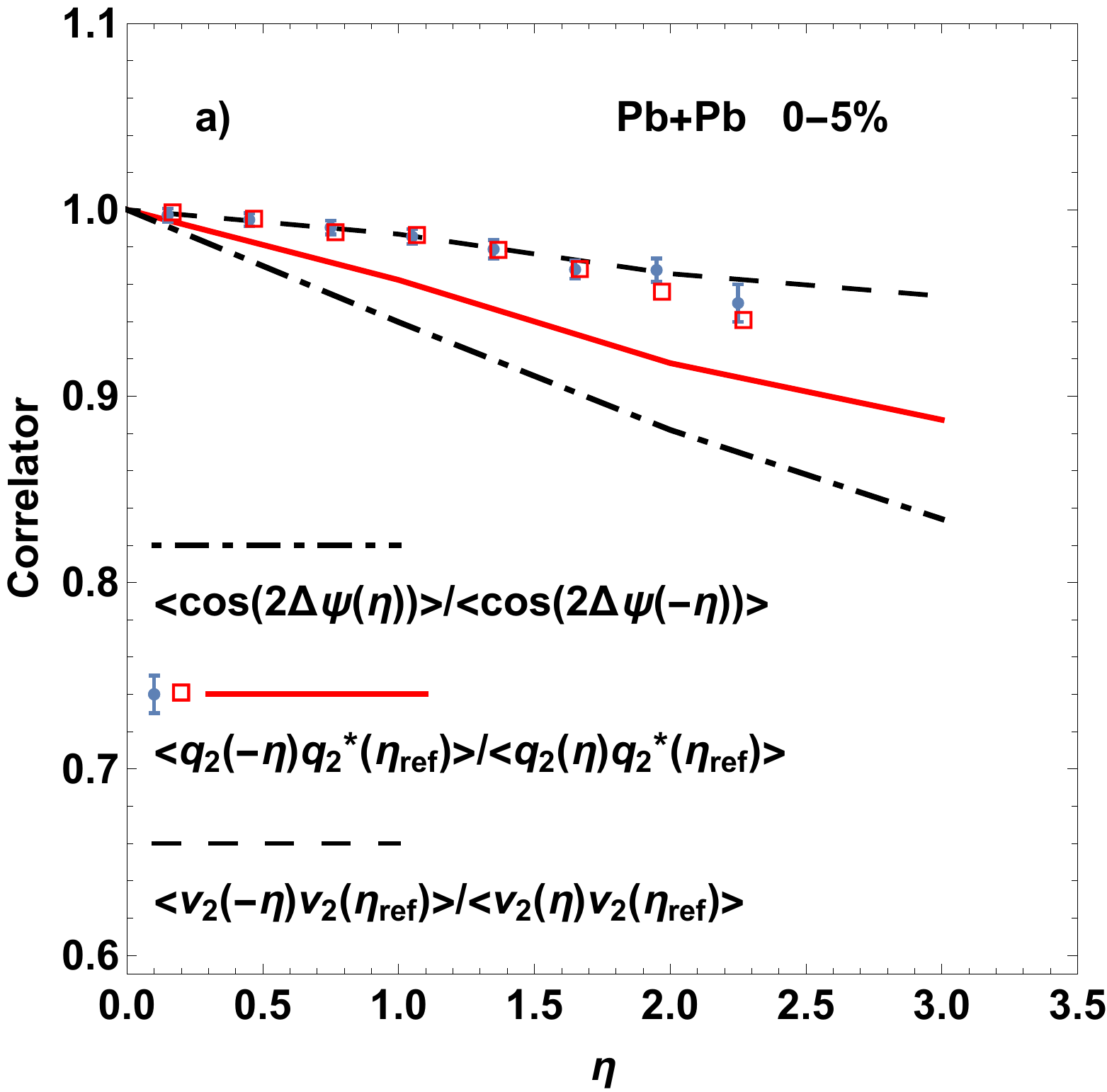}~~~~~~~~~~~
 \includegraphics[width=0.4\textwidth]{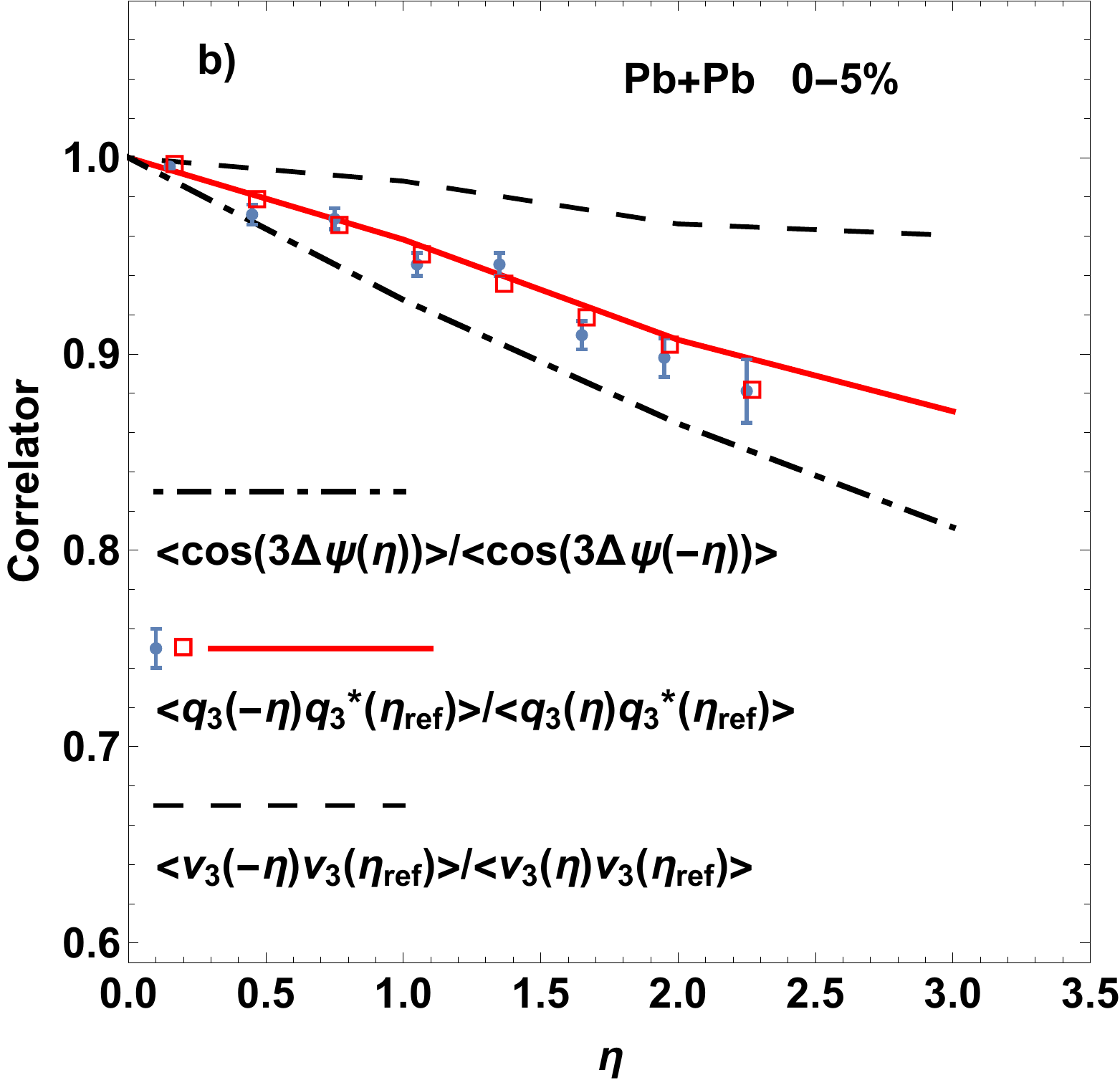}
 \caption{The torque angle decorrelation (dash-dotted line), the magnitude decorrelation (dashed line) 
 and the torque and magnitude decorrelation combined (solid line) from  the hydrodynamic model
 for the elliptic (left panel) and triangular (right panel) flow. The points denote the data from 
 ATLAS and CMS Collaborations \cite{Khachatryan:2015oea,Aaboud:2017tql}  (from \cite{Bozek:2017qir}).}
 \label{fig:angmag}
 \end{center}
\end{figure}

\begin{figure}[h]
 \begin{center}
 \includegraphics[width=0.4\textwidth]{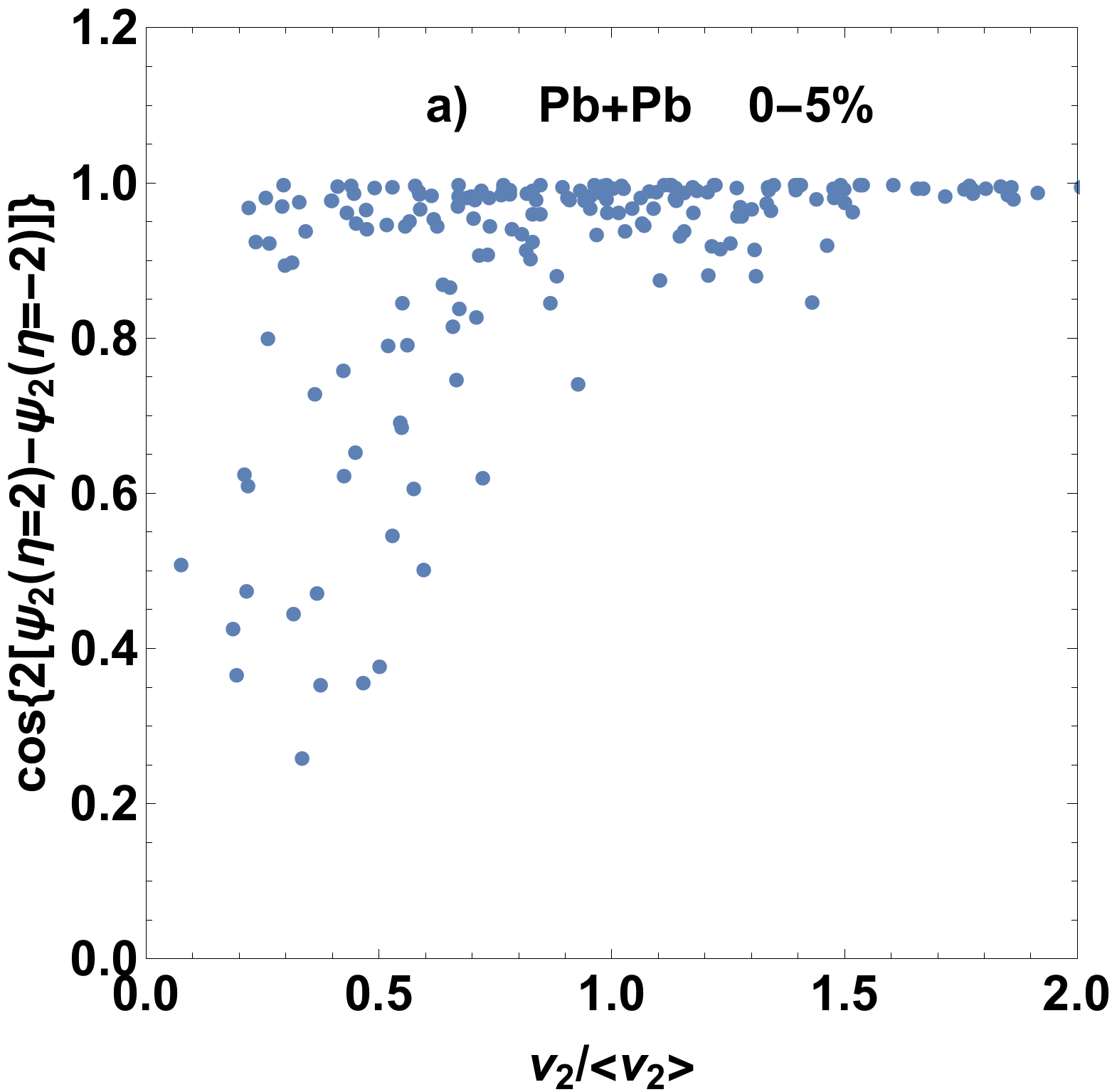}~~~~~~~~~~~
 \includegraphics[width=0.4\textwidth]{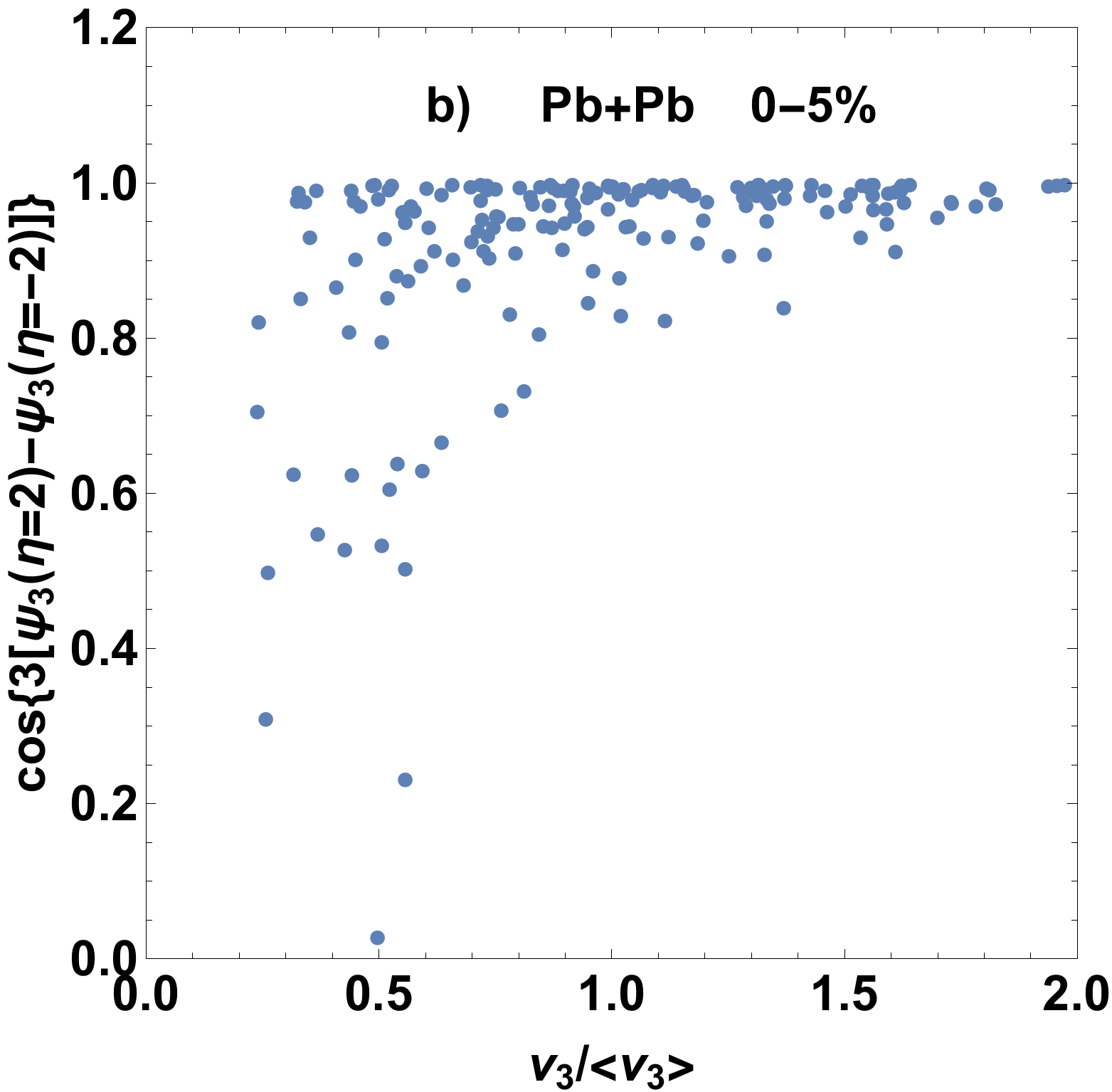}
 \caption{Scatter plot for the event by event torque angle decorrelation and the flow magnitude from 
 the hydrodynamic model for the elliptic (left panel) and triangular (right panel) flow  (from \cite{Bozek:2017qir}).}
 \label{fig:rvk}
 \end{center}
\end{figure}

The decorrelation of the harmonic flow magnitude and angle can be studied 
separately in the hydrodynamic model 
\cite{Bozek:2017qir}. In the model, the angle and the magnitude of the collective flow can be 
estimated in each event. Besides the experimental observable $r_{n,1}(\eta)$ (Eq. \ref{eq:dvn}) one can calculate the torque angle decorrelation
\begin{equation}
\frac{\langle \cos\left(n (\Psi_n(-\eta) -\Psi_n(\eta_{ref})) \right)\rangle}
{\langle \cos\left(n (\Psi_n(\eta) -\Psi_n(\eta_{ref})) \right)\rangle}
\label{eq:tw}
\end{equation}
and the magnitude decorrelation 
\begin{equation}
\frac{\langle v_n(-\eta)v_n(\eta_{\rm ref})\rangle}
{\langle  v_n(\eta)v_n(\eta_{\rm ref}) \rangle}
\end{equation}
directly. The results shown in Fig.~\ref{fig:angmag} are quite surprising; one observes a sort of inverted hierarchy. The torque angle  decorrelation 
is the largest, {\it larger} than the magnitude and torque angle decorrelation combined. Interestingly, 
a very similar result has been obtained recently in a hybrid model using hydrodynamics with the AMPT initial conditions \cite{Wu:2018cpc}.

This inverted hierarchy can be explained by the fact that the torque angle decorrelation (Eq. \ref{eq:tw}) is effectively 
weighted by a different power of the flow magnitude ($v_n^0$) than the correlator $r_{n,1}(\eta)$ ($v_n^2$).
It turns out that the magnitude of the flow coefficient $v_n$ is strongly anticorrelated with the magnitude of the torque angle. In events
with a large overall harmonic flow, the torque angle is smaller ($\cos(\Delta \Psi_n)$  close to $1$ in Fig. \ref{fig:rvk}). In events 
with a small magnitude of harmonic flow, the torque angle distribution is wider ($\cos(\Delta \Psi_n)$  deviates from
 $1$ in Fig. \ref{fig:rvk}). The correlation visible in the scatter plots explains why the average $\langle \cos(\Delta \Psi_n) \rangle$ is closer to $1$ 
 when effectively weighted with a higher power of $v_n$,.

\begin{figure}
 \begin{center}
 \includegraphics[width=0.4\textwidth]{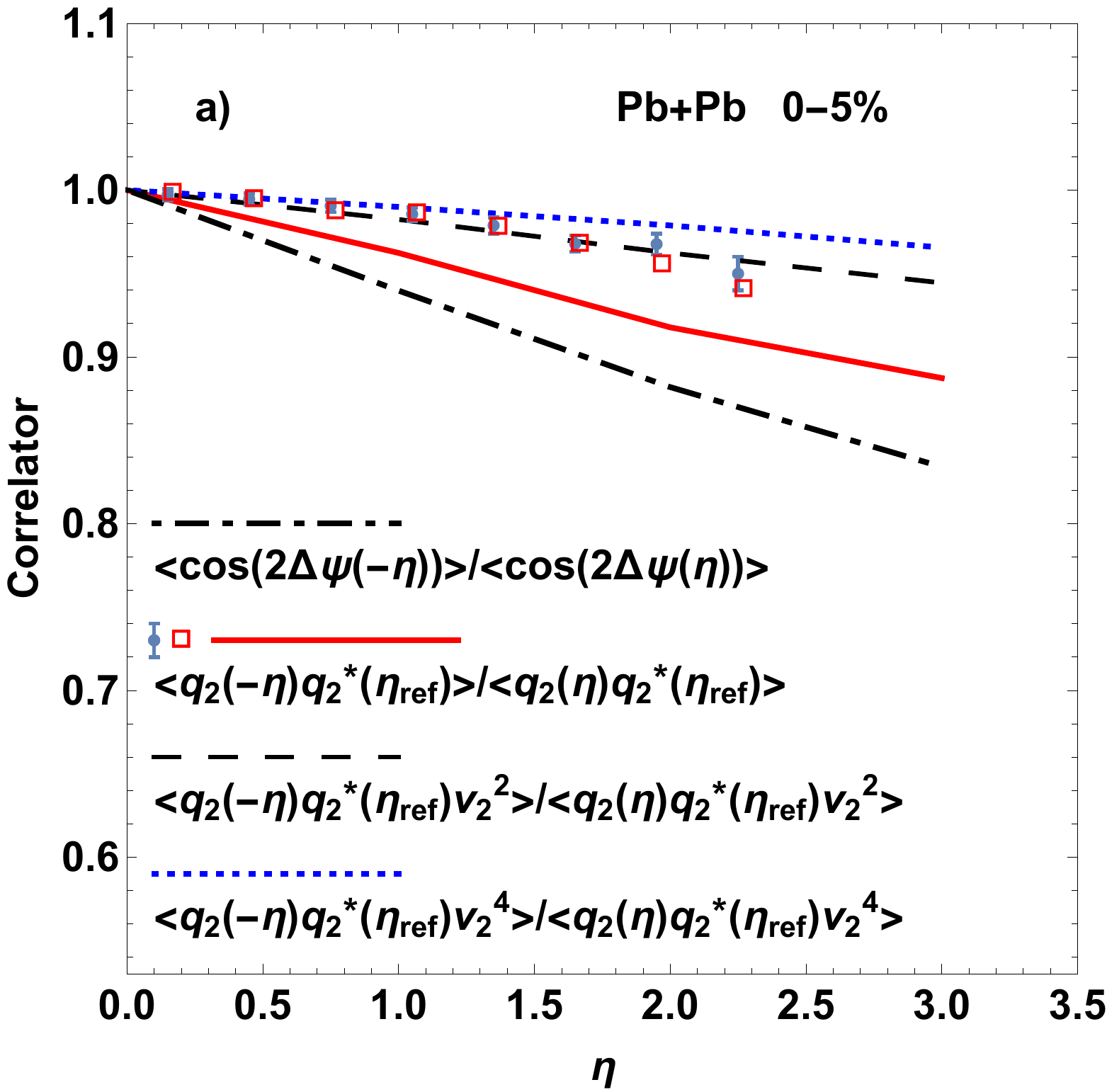}~~~~~~~~~~~
 \includegraphics[width=0.4\textwidth]{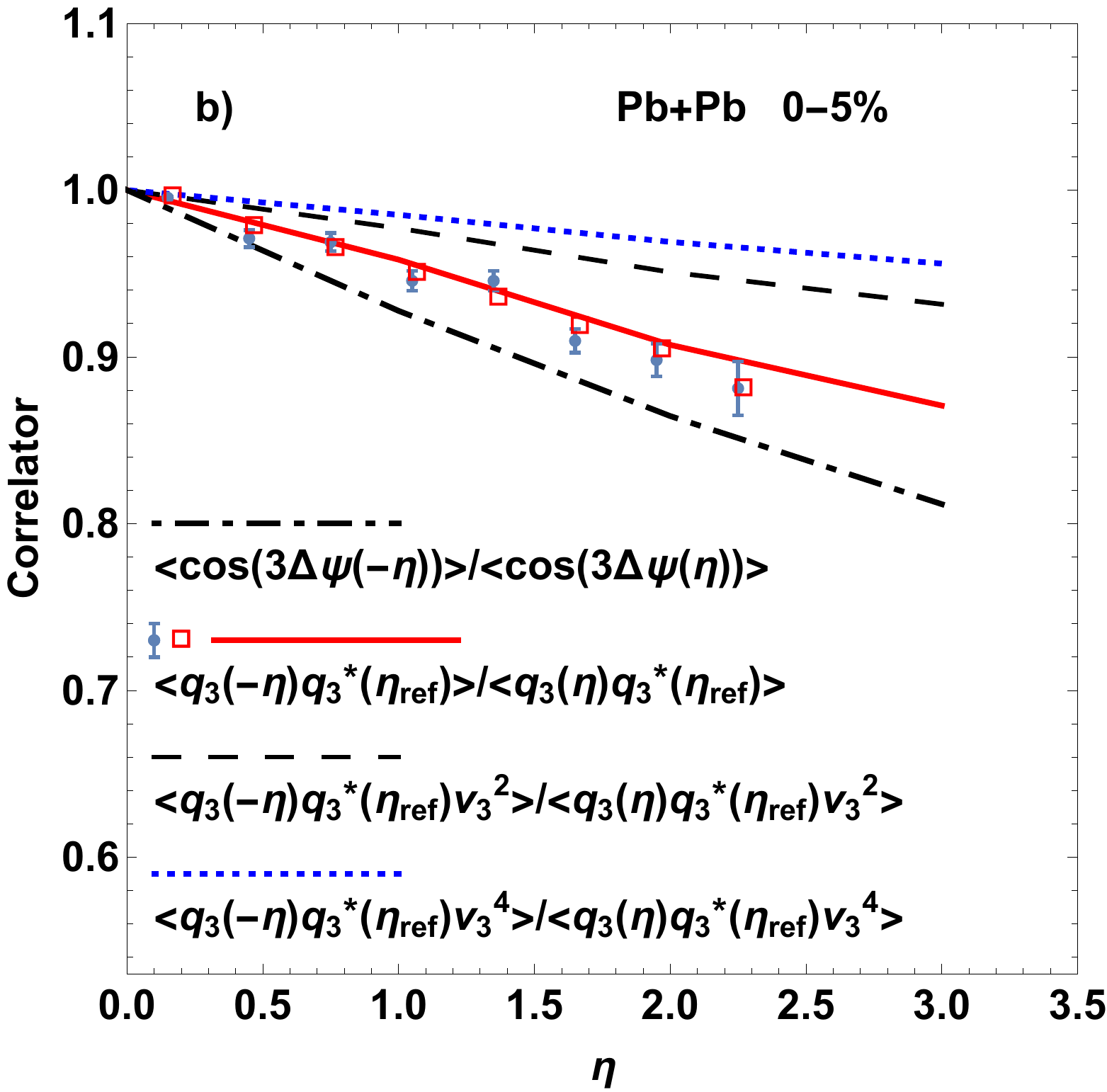}
 \caption{The torque angle decorrelation (dash-dotted line) and the harmonic flow correlator  weighted 
 with different powers of $v_n$  (dotted, dashed and solid lines) from the  hydrodynamic model 
 for the elliptic (left panel) and triangular (right panel) flow.  The points denote the data from ATLAS and CMS Collaborations 
 \cite{Khachatryan:2015oea,Aaboud:2017tql}  (from \cite{Bozek:2017qir}).}
 \label{fig:rvk}
 \end{center}
\end{figure}

Experimentally,  event by event correlation between the flow magnitude and the
torque angle can be demonstrated using  a series of correlators weighted with different powers of $v_n^{2k}$,
\begin{equation}
r_{n|n;1}^{n;2k}(\eta)=\frac{\langle v_n^{2k}(0) q_n(-\eta) q_n^\star(\eta_{ref}) \rangle}{ \langle v_n^{2k}(0) q_n(\eta) q_n^\star(\eta_{ref}) \rangle} ,
\end{equation}
for $k=0,1,2,\dots$. The case $k=0$ corresponds to the standard correlator $r_{n|n;1}(\eta)$. The model predicts a
weaker decorrelation (correlator closer to $1$) when weighting with higher powers of $v_n$ (Fig. \ref{fig:rvk}). It would be interesting to measure this correlation in experiment.

\vskip 3mm
\noindent
{\bf Acknowledgments}\\
Research  supported by the  AGH UST statutory funds, by the National
Science Centre grant 2015/17/B/ST2/00101 (PB) and grant 2015/19/B/ST2/00937 
(WB).

\bibliographystyle{elsarticle-num}
\bibliography{../../../hydr.bib}







\end{document}